\documentclass[
    ,final            
  ]
  {aipproc}

\layoutstyle{6x9}

\begin{document}

\title{Experimental Evidence for Efimov Quantum States}


\author{H.-C. N\"agerl}{
  address={Institut f\"ur Experimentalphysik, University of Innsbruck, Innsbruck, Austria}
}

\author{T. Kraemer}{
  address={Institut f\"ur Experimentalphysik, University of Innsbruck, Innsbruck, Austria}
}

\author{M. Mark}{
  address={Institut f\"ur Experimentalphysik, University of Innsbruck, Innsbruck, Austria}
}

\author{P. Waldburger}{
  address={Institut f\"ur Experimentalphysik, University of Innsbruck, Innsbruck, Austria}
}

\author{J. G. Danzl}{
  address={Institut f\"ur Experimentalphysik, University of Innsbruck, Innsbruck, Austria}
}

\author{B. Engeser}{
  address={Institut f\"ur Experimentalphysik, University of Innsbruck, Innsbruck, Austria}
}

\author{A. D. Lange}{
  address={Institut f\"ur Experimentalphysik, University of Innsbruck, Innsbruck, Austria}
}

\author{K. Pilch}{
  address={Institut f\"ur Experimentalphysik, University of Innsbruck, Innsbruck, Austria}
}

\author{A. Jaakkola}{
  address={Institut f\"ur Experimentalphysik, University of Innsbruck, Innsbruck, Austria}
}

\author{C. Chin}{
  address={James Franck Institute, Physics Department of the University of Chicago, 5640 S. Ellis Ave. Chicago, Illinois 60637, USA}
}

\author{R. Grimm}{
  address={Institut f\"ur Experimentalphysik, University of Innsbruck, Innsbruck, Austria}
  ,altaddress={Institut f\"ur Quantenoptik und Quanteninformation, \"Osterreichische Akademie der Wissenschaften, Innsbruck, Austria} 
}

\begin{abstract}
Three interacting particles form a system which is well known for
its complex physical behavior. A landmark theoretical result in
few-body quantum physics is Efimov's prediction of a universal set
of weakly bound trimer states appearing for three identical bosons
with a resonant two-body interaction \cite{Efimov70, Efimov71}.
Surprisingly, these states even exist in the absence of a
corresponding two-body bound state and their precise nature is
largely independent of the particular type of the two-body
interaction potential. Efimov's scenario has attracted great
interest in many areas of physics; an experimental test however
has not been achieved. We report the observation of an Efimov
resonance in an ultracold thermal gas of cesium atoms
\cite{Kraemer06}. The resonance occurs in the range of large
negative two-body scattering lengths and arises from the coupling
of three free atoms to an Efimov trimer. We observe its signature
as a giant three-body recombination loss when the strength of the
two-body interaction is varied near a Feshbach resonance. This
resonance develops into a continuum resonance at non-zero
collision energies, and we observe a shift of the resonance position as a
function of temperature. We also report on a minimum in the
recombination loss for positive scattering lengths, indicating
destructive interference of decay pathways. Our results confirm
central theoretical predictions of Efimov physics and represent a
starting point from which to explore the universal properties of
resonantly interacting few-body systems.
\end{abstract}

\maketitle



Efimov's treatment of three identical bosons \cite{Efimov70,
Efimov71} is closely linked to the concept of universality
\cite{Braaten04} in systems with a resonant two-body interaction,
where the s-wave scattering length $a$ fully characterizes the
two-body physics. When $|a|$ greatly exceeds the characteristic
range $\ell$ of the two-body interaction potential, details of the
short-range interaction become irrelevant because of the
long-range nature of the wave function. Universality then leads to
a generic behavior in three-body physics, reflected in the energy
spectrum of weakly bound Efimov trimer states. Up to now, in spite
of their great fundamental importance, these states could not be
observed experimentally. An observation in the realm of nuclear
physics, as originally proposed by Efimov, is hampered by the
presence of the Coulomb interaction, and only two-neutron halo
systems with a spinless core are likely to feature Efimov states
\cite{Jensen04}. In molecular physics, the helium trimer
\cite{Schoellkopf94} is predicted to have an excited state with
Efimov character \cite{Lim77}. The existence of this state could
so far not be confirmed \cite{Bruehl05}. A different approach to
experimentally study the physics of Efimov states is based on the
unique properties of ultracold atomic quantum gases. In such systems \cite{ultracoldmatter}
an unprecedented level of
control is possible to study interacting quantum systems. The ultra-low
collision energies allow to explore the zero-energy quantum limit
and two-body interactions can be precisely tuned based on Feshbach
resonances \cite{Tiesinga93,Inouye98}.
\begin{figure}
  \includegraphics[height=.25\textheight]{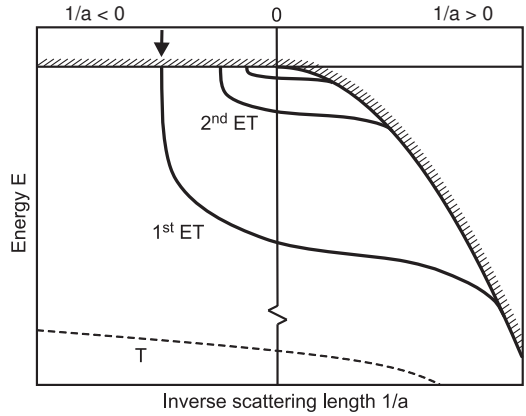}
  \caption{Efimov's scenario: Appearance of an infinite series of
weakly bound Efimov trimer states (ET) for resonant two-body
interaction. The binding energy is plotted as a function of the
inverse two-body scattering length $1/a$. The hatching
indicates the scattering continuum for three atoms (for
$a<0$) and for an atom and a dimer (for $a>0$). The arrow marks
the intersection of the first Efimov trimer with the three-atom
threshold. To illustrate the series of Efimov states, we have
artificially reduced the universal scaling factor from $22.7$ to
$2$. For comparison, the dashed line (T) indicates the presence of a tightly bound
non-Efimov trimer \cite{Thomas35} which does not interact with the
scattering continuum.} \label{Efimov}
\end{figure}

Efimov's scenario \cite{Efimov70,Efimov71,Braaten04} can be
illustrated by the energy spectrum of the three-body system as a
function of the inverse scattering length $1/a$
(Fig.~\ref{Efimov}). Let us first consider the well-known weakly
bound dimer state, which only exists for large positive $a$. In
the resonance regime, its binding energy is given by the universal
expression $E_b =-\hbar^2/(ma^2)$, where $m$ is the atomic mass
and $\hbar$ is Planck's constant divided by $2\pi$. In
Fig.~\ref{Efimov}, where the resonance limit corresponds to $1/a
\rightarrow 0$, the dimer energy $E_b$ is represented by a
parabola for $a>0$. If we now add one more atom with zero energy,
a natural continuum threshold for the bound three-body system
(hatching in Fig.~\ref{Efimov}) is given by the
three-atom threshold ($E =0$) for negative $a$ and by the
dimer-atom threshold ($E_b$) for positive $a$. Energy states below
the continuum threshold are necessarily three-body bound states.
When $1/a$ approaches the resonance from the negative-$a$ side, a
first Efimov trimer state appears in a range where a weakly bound
two-body state does not exist. When passing through the resonance
the state connects to the positive-$a$ side, where it finally
intersects with the dimer-atom threshold. An infinite series of
such Efimov states is found when scattering lengths are increased and binding energies
are decreased in powers of universal scaling factors $e^{\pi/s_0}
\approx 22.7$ and $e^{-2\pi/s_0} \approx 1/515$ ($s_0 = 1.00624$),
respectively \cite{Efimov70,Efimov71,Braaten04}.


Resonant scattering phenomena arise as a natural consequence of
Efimov's scenario \cite{Efimov79}. When an Efimov state intersects
with the continuum threshold at negative scattering lengths $a$,
three free atoms in the ultracold limit resonantly couple to a
trimer. This results in a triatomic Efimov resonance
\cite{Esry99,Braaten01}. At finite collision energies, the
phenomenon evolves into a triatomic continuum resonance
\cite{Bringas04}. Another type of Efimov resonance
\cite{Nielsen02} is found at positive values of $a$ for collisions
between a free atom and a dimer, when Efimov states intersect with
the dimer-atom threshold. While the latter type of Efimov
resonance corresponds to Feshbach resonances in collisions
between atoms and dimers \cite{Nielsen02}, triatomic Efimov
resonances can be interpreted as a three-body generalization to
Feshbach resonances \cite{Stoll05}.

Striking manifestations of Efimov physics have been predicted for
three-body recombination processes in ultracold gases with tunable
two-body interactions \cite{Braaten04,Esry99,Braaten01,
Nielsen99,Bedaque00,DIncao04}. Three-body recombination leads to
losses from a trapped gas with a rate proportional to the third
power of the atomic number density. These losses are commonly
described \cite{Weber03b} in terms of a loss rate coefficient
$L_3$. In the resonant case ($|a| \gg \ell$), it is convenient to
express this coefficient in the form $L_3 = 3 \, C(a) \hbar a^4
/m$, separating a general $a^4$-scaling \cite{Weber03b,Fedichev96}
from an additional dependence \cite{Esry99,Braaten01,Bedaque00}
$C(a)$. Efimov physics is reflected in a logarithmically periodic
behavior $C(22.7a) = C(a)$, corresponding to the scaling of the
infinite series of weakly bound trimer states. For negative
scattering lengths, the resonant coupling of three atoms to an
Efimov state opens up fast decay channels into deeply bound dimer
states plus a free atom. Triatomic Efimov resonances thus show up
in giant recombination loss. This striking phenomenon was first
identified in numerical solutions to the adiabatic hyperspherical
approximation of the three-body Schr\"odinger equation assuming
simple model potentials and interpreted in terms of tunneling
through a potential barrier in the three-body entrance channel
\cite{Esry99}. A different theoretical approach \cite{Braaten04,
Braaten01}, based on effective field theory, provides the analytic
expression $C(a) = 4590 \sinh{(2\eta_-)} / \left( \sin^2{[s_0
\ln{(|a|/a_-)} ]} + \sinh^2{\eta_-} \right)$. The free parameter
$a_-$ for the resonance positions at $a_-$, $22.7\,a_-$, $...$
depends on the short-range part of the three-body interaction and
is thus not determined in the frame of the universal long-range
theory. As a second free parameter, the dimensionless quantity
$\eta_-$ describes the unknown decay rate of Efimov states into
deeply bound dimer states plus a free atom. It thus characterizes
the resonance width.


Our measurements are based on the magnetically tunable interaction
properties of Cs atoms \cite{Chin04} in the lowest internal state
(quantum numbers $F=3$ for the total spin and $m_F=3$ for its
projection). By applying fields between $0$ and $150$\,G, we can
vary the s-wave scattering length $a$ in a range between
$-2500\,a_0$ to $1600\,a_0$, where $a_0$ is Bohr's radius. The
dependence can in general be well approximated by the fit formula
\begin{equation}
a(B)/a_0 = \left(1722+1.52\,B/{\rm G}\right) \left(1 -
\frac{28.72}{B/{\rm G}+11.74}\right), \nonumber
\end{equation}
except for a few narrow Feshbach resonances \cite{Chin04}. The
smooth variation of the scattering length in the low-field region
results from a broad Feshbach resonance centered at about $-12\,$G
(equivalent to $+12\,$G in the state $F=3$, $m_F=-3$). In all our
measurements we excluded the magnetic field regions where the
narrow Feshbach resonances influence the scattering behavior
through coupling to other molecular potentials. Accurate
three-body loss measurements are facilitated by the fact that
inelastic two-body loss is energetically forbidden
\cite{Weber03b}. The characteristic range of the two-body
potential is given by the van der Waals length \cite{vdWlength},
which for Cs is $\ell \approx 100\,a_0$. This leaves us with
enough room to study the universal regime requiring $|a| \gg
\ell$. For negative $a$, a maximum value of $25$ is attainable for
$|a|/\ell$. Efimov's estimate $\frac{1}{\pi}\,\ln(|a|/\ell)$ for
the number of weakly bound trimer states \cite{Efimov71} suggests
the presence of one triatomic Efimov resonance in the accessible range of
negative scattering lengths.

All measurements were performed with trapped thermal samples of Cs
atoms at temperatures $T$ ranging from 10 to 250\,nK. We used two
different experimental setups, which have been described elsewhere
\cite{Weber03a,Kraemer04,Rychtarik04}. In setup A we first
produced an essentially pure Bose-Einstein condensate (BEC) with
up to 250,000 atoms in a far-detuned crossed optical dipole trap
generated by two 1060-nm Yb-doped fiber laser beams
\cite{Kraemer04}. We then ramped the magnetic field to 16.2\,G
where the scattering length is negative with a value of
$-50\,a_0$, thus inducing a collapse of the condensate
\cite{Donley01}. After an equilibration time of 1\,s we were left
with a thermal sample at typically $T = 10$\,nK containing up to
20,000 atoms at peak densities ranging from $n_0 = 3 \times
10^{11}$\,cm$^{-3}$ to $3 \times 10^{12}$\,cm$^{-3}$.
Alternatively, we interrupted the evaporation process before
condensation to produce thermal samples at $T \approx 200$\,nK in
a crossed dipole trap generated by one of the 1060-nm beams and a
$10.6$-$\mu$m CO$_2$-laser beam. After recompression of the trap
this produced typical densities of  $n_0 = 5 \times
10^{13}$\,cm$^{-3}$. To determine the three-body loss rates in
this setup we recorded the time evolution of the atom number $N$
and the temperature $T$. A detailed description of this procedure
has been given in Ref.~\cite{Weber03b}. In brief, the process of
three-body recombination not only leads to a loss of atoms, but
also induces ``anti-evaporation'' and recombination heating. The
first effect is present at any value of the scattering length $a$.
The second effect occurs for positive values of $a$ when the
recombination products remain trapped. Atom loss and temperature
rise are modelled by a set of two coupled non-linear differential
equations. We used numerical solutions to this set of equations to
fit our experimental data. From these fits together with
measurements of the trapping parameters we obtained the rate
coefficient $L_3$. Alternatively, we simply measured the loss
fraction after some fixed storage time as a function of magnetic field,
i.e. atomic scattering length. For this the storage time was
200\,ms at initial densities of $n_0 = 6 \times
10^{13}$\,cm$^{-3}$. In setup B we used an optical surface trap
\cite{Rychtarik04} in which we prepared a thermal sample of 10,000
atoms at $T \approx 250$\,nK via forced evaporation at a density
of $n_0 = 1.0 \times 10^{12}$\,cm$^{-3}$. The dipole trap was
formed by a repulsive evanescent laser wave on top of a horizontal
glass prism in combination with a single horizontally confining
1060-nm laser beam propagating along the vertical direction. In
this setup we measured three-body loss rates by recording the loss
as a function of time at sufficiently short decay times for which
heating is negligible. To determine the position of maximum
three-body loss we again chose the simpler procedure of measuring
the loss fraction after some fixed storage time as a function of
magnetic field.
\begin{figure}
\includegraphics[height=.25\textheight]{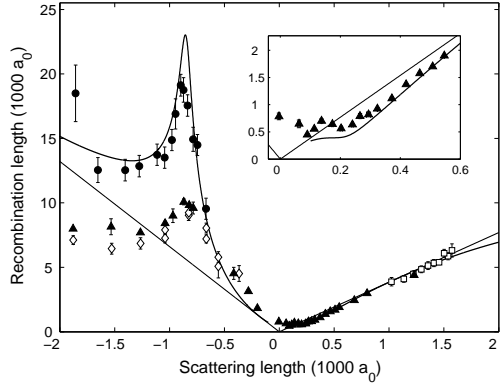}
\caption{Observation of the Efimov resonance in measurements of
three-body recombination. The recombination length $\rho_3 \propto
L_3^{1/4}$ is plotted as a function of the scattering length $a$.
The dots and the filled triangles show the experimental data from
setup A for initial temperatures around 10\,nK and 200\,nK,
respectively. The open diamonds are from setup B at temperatures
of 250\,nK. The open squares are previous data \cite{Weber03b} at
initial temperatures between 250 and 450\,nK. The solid curve
represents the analytic model from effective field theory
\cite{Braaten04} with $a_- = -850\,a_0$, $a_+ = 1060\,a_0$, and
$\eta_- = \eta_+ = 0.06$. The straight lines result from setting
the $\sin^2$ and $\cos^2$-terms in the analytic theory to $1$,
which gives a lower recombination limit for $a<0$ and an upper
limit for $a>0$. The inset shows an expanded view for small
positive scattering lengths with a minimum for $C(a) \propto
(\rho_3/a)^4$ near $210\,a_0$. The displayed error bars refer to
statistical uncertainties only. Uncertainties in the determination
of the atomic number densities may lead to additional calibration
errors for $\rho_3$ of up to 20\%.} \label{Rec_length}
\end{figure}


Our experimental results (Fig.~\ref{Rec_length}) indeed show a
giant loss feature marking the expected resonance. The Efimov
resonance is centered at 7.5\,G. We present our data in terms of a
recombination length \cite{Esry99} $\rho_3 =
[2m/(\sqrt{3}\hbar)\,L_3]^{1/4}$, which leads to the simple
relation $\rho_3/a = 1.36\,C^{1/4}$. Note that the general
$a^4$-scaling corresponds to a linear behavior in $\rho_3(a)$
(straight lines in Fig.~\ref{Rec_length}). A fit of the analytic
theory \cite{Braaten04,Braaten01} to our experimental data taken
for negative $a$ at temperatures $T\approx10$\,nK shows a
remarkable agreement and determines the resonance position to
$a_-=-850(20)\,a_0$ and the decay parameter to $\eta_-=0.06(1)$.
The pronounced resonance behavior with a small value for the decay
parameter ($\eta_- \ll 1$) demonstrates a sufficiently long
lifetime of Efimov trimers to allow their observation as distinct
quantum states.

All the results discussed so far are valid in the zero-energy
collision limit of sufficiently low temperatures. For ultralow but
non-zero temperatures the recombination length is unitarity
limited \cite{DIncao04} to $5.2 \, \hbar \, (m k_B T)^{-1/2}$. For
$T=10$\,nK this limit corresponds to about $60,000\,a_0$ and our
sample is thus cold enough to justify the zero-temperature limit.
For 250\,nK, however, unitarity limits the recombination length to
about $12,000\,a_0$. The Efimov resonance is still visible at
temperatures of 200 and 250\,nK (filled triangles and open
diamonds in Fig.~\ref{Rec_length}). The slight shift to lower
values of $|a|$ suggests the evolution of the zero-energy Efimov
resonance into a triatomic continuum resonance \cite{Bringas04}.
In further experiments at higher temperatures (data not shown) we
observed the resonance to disappear above $\sim$500\,nK.

\begin{figure}
\includegraphics[height=.2\textheight]{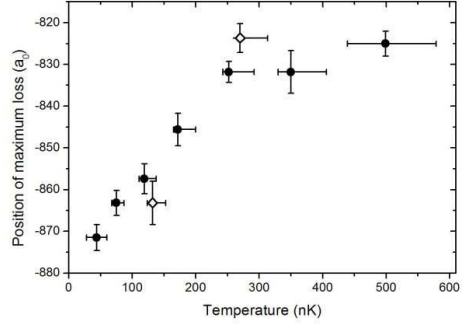}
\caption{Position of maximum three-body loss for different
temperatures of the atomic sample (filled circles) as measured in
setup B. Two data points were measured after recompression of the
optical trap (open diamonds). For lower temperatures, the position of peak loss
shifts to higher absolute values of the scattering length.} \label{shift}
\end{figure}

To quantify the shift of the resonance position \cite{Engeser06} we have determined
the magnetic field value of maximum loss for temperatures in the
range from $500$\,nK down to $44$\,nK in setup B. We observe the
peak position to shift from $7.7$\,G at high temperatures to
$7.4$\,G for the lowest temperature. Fig.~\ref{shift} presents
the data after conversion of the magnetic field values to the
corresponding scattering lengths. The loss peak shifts from
$-825\,a_0$ at $500$\,nK down to $-872\,a_0$ for the coldest samples. Two data
points have been measured after deep evaporation followed by a
recompression ramp that increased the trapping potential by a
factor of three within $900$\,ms. The results are consistent with
the measurements in uncompressed traps and show that plain
evaporation is absent. The vertical error bars indicate the
statistical uncertainty from the parabolic fitting procedure. The
horizontal error bars take into account the uncertainty of the
temperature measurements and the expected rise in temperature that
accompanies the loss \cite{Weber03b}.

For positive scattering lengths, we found three-body losses to be
typically much weaker than for negative values. Our measurements
are consistent with a maximum recombination loss of $C(a)\approx
70$, or equivalently $\rho_3 \approx 3.9\,a$, as predicted by
different theories \cite{Esry99,Nielsen99,Bedaque00} (straight
line for $a>0$ in Fig.~\ref{Rec_length}). For $a$ below $600\,a_0$
the measured recombination length significantly drops below this
upper limit (inset in Fig.~\ref{Rec_length}). The analytic expression from effective field
theory \cite{Braaten04,Bedaque00} for $a>0$ reads $C(a) = 67.1
\,e^{-2\eta_+} \, \left( \cos^2[s_0\ln(a/a_+)] + \sinh^2\eta_+
\right) + 16.8\, (1-e^{-4\eta_+})$ with the two free parameters
$a_+$ and $\eta_+$. The first term describes recombination into
the weakly bound dimer state with an oscillatory behavior due to
an interference effect between two different pathways
\cite{Esry99,Nielsen99}. The second term results from decay into
deeply bound states.
We use this expression to fit our data points with $a>5\,\ell
\approx 500a_0$. This somewhat arbitrary condition is introduced
as a reasonable choice to satisfy $a \gg \ell$ for the validity of
the universal theory. The fit is quite insensitive to the value of
the decay parameter and yields $\eta_+ < 0.2$. This result is
consistent with the theoretical assumption \cite{Braaten01} of the
same value for the decay parameter for positive and negative $a$,
which in our case is $\eta_+ = \eta_- =0.06$. For maximum loss, we
obtain $a_+ = 1060(70)\,a_0$. According to theory
\cite{Braaten04}, the trimer state hits the dimer-atom threshold
at $a = 1.1\,a_+ \approx 1170\,a_0$. The logarithmic periodicity
of the Efimov scenario suggests adjacent loss minima to occur at
$\sqrt{22.7} \times1060\,a_0 \approx 5000\,a_0$ and at
$1060\,a_0/\sqrt{22.7} \approx 220\,a_0$. While the former value
is out of our accessible range, the latter value ($a\approx
2\ell$) is too small to strictly justify universal behavior in the
resonance limit ($a \gg \ell$). Nevertheless, our experimental
results (inset to Fig.~\ref{Rec_length}) indicate a minimum at $a
\approx 210\,a_0$ and the analytic expression for $C(a)$ is found
to describe our data quite well down to this minimum.
\begin{figure}
\includegraphics[height=.2\textheight]{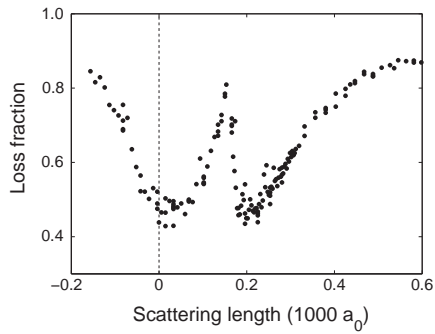}
\caption{Atom loss for small scattering lengths as measured in setup A.
Besides a minimum
near zero scattering length, we identify a minimum of
recombination loss at $\sim$$210\,a_0$, which can be attributed to
a predicted destructive interference effect
\cite{Esry99, Nielsen99,Bedaque00}.} \label{loss_fraction}
\end{figure}

The occurrence of the interference minimum in three-body loss is
demonstrated more clearly in another set of experiments where we simply measured the loss of
atoms after a fixed storage time in the optical trap (Fig.~\ref{loss_fraction}). This minimum
is located at $a=210(10)\,a_0$ in addition to a second minimum
close to zero scattering length. We point out that the existence
of the minimum at $210\,a_0$ is very advantageous for efficient
evaporative cooling of Cs as it combines a large scattering cross
section with very low loss. Inadvertently, we have already
benefitted of this loss minimum for the optimized production of a
Bose-Einstein condensate of Cs \cite{Kraemer04}.

The comparison of our experimental results to available three-body
theory shows remarkable agreement, although the collision physics
of Cs is in general a very complicated multi-channel scattering
problem. We believe that the particular nature of the broad,
``open-channel dominated'' Feshbach resonance \cite{Koehler06}
that underlies the tunability of our system plays a crucial role.
For such a resonance, the two-body scattering problem can be
reduced to an effective single-channel model. It is a very
interesting question to what degree this great simplification of
the two-body physics extends to the three-body problem. In particular, we
raise the question how the regions of positive and negative
scattering lengths are connected in our experiment, where $a$ is
changed through a zero crossing, i.e.\ through a non-universal
region, and not across the universal resonance region. In our
case, there is no obvious connection between the Efimov state that
leads to the observed resonance for $a<0$ and the states
responsible for the behavior for $a>0$. In our analysis of the
experimental data, we have thus independently fitted the data sets
for negative and positive $a$. Nevertheless, the resulting values
for the two independent fit parameters $a_-$ and $a_+$ suggest a
connection: For the ratio $a_+/|a_-|$ our experiment yields
1.25(9), whereas universal theory \cite{Braaten04} predicts
0.96(3). These numbers are quite close in view of the Efimov
factor of 22.7. If not an accidental coincidence, we speculate
that the apparent relation between $a_+$ and $a_-$ may be a
further consequence of universality in a system where the resonant
two-body interaction can be modelled in terms of a single
scattering channel. In general, the multi-channel nature of
three-body collisions near Feshbach resonances
\cite{Kartavtsev02,Petrov04} leads to further interesting
questions, like e.g.\ possible resonance effects beyond the Efimov
scenario. Advances in three-body theory are necessary to
answer these questions and to provide a complete interpretation of
our present observations.

In the past few years, applications of Feshbach resonances in
ultracold gases and the resulting possibility to create dimer
states have set the stage for many new developments in matter-wave
quantum physics. The observation of an Efimov resonance now
confirms the existence of weakly bound trimer states and opens up
new vistas \cite{Stoll05,Braaten03} to experimentally explore the
intriguing physics of few-body quantum systems.


\begin{theacknowledgments}
We thank E. Braaten, C. Greene, B. Esry, H. Hammer, and T.
K{\"o}hler for many stimulating and fruitful discussions and E.
Kneringer for support regarding the data analysis. We acknowledge
support by the Austrian Science Fund (FWF) within
Spezialforschungsbereich 15 and within the Lise Meitner program,
and by the European Union in the frame of the TMR networks ``Cold
Molecules'' and ``FASTNet''. M.M. is supported within the
Doktorandenprogramm of the Austrian Academy of Sciences.
\end{theacknowledgments}


\end{document}